\documentclass[aps,prb,amsfonts,amssymb,twocolumn,amsmath,preprintnumbers,floatfix,superscriptaddress]{revtex4-2}%
\usepackage{appendix}
\usepackage{siunitx}
\usepackage{graphicx}
\usepackage{mathrsfs}
\usepackage{bm}
\usepackage{amsmath}
\usepackage{dcolumn}
\usepackage{epstopdf}
\usepackage{dsfont}
\usepackage{amssymb}
\usepackage{tabularx}
\usepackage{longtable}
\usepackage{array}
\usepackage{float}
\usepackage{color}
\usepackage{epstopdf}
\usepackage{mathrsfs}
\usepackage{multirow}
\usepackage[T1]{fontenc}
\usepackage[colorlinks,linkcolor=blue,anchorcolor=blue,citecolor=blue,urlcolor=blue]{hyperref}

\renewcommand{\Re}{\operatorname{Re}}
\renewcommand{\Im}{\operatorname{Im}}
\begin{document}

\title{Disorder induced time-reversal-odd nonlinear spin and orbital Hall effects}

\author{Ruda Guo}
\affiliation{Center for Advanced Quantum Studies and School of Physics and Astronomy, Beijing Normal University, Beijing 100875, China}
\affiliation{State Key Laboratory of Surface Physics and Interdisciplinary Center for Theoretical Physics and Information Sciences, Fudan University, Shanghai 200433, China}

\author{Yi Liu}
\email{yiliu42@shu.edu.cn}
\affiliation{Institute for Quantum Science and Technology, Shanghai University, Shanghai 200444, China}
\affiliation{Department of Physics, Shanghai University, Shanghai 200444, China}

\author{Cong Xiao}
\email{congxiao@fudan.edu.cn}
\affiliation{State Key Laboratory of Surface Physics and Interdisciplinary Center for Theoretical Physics and Information Sciences, Fudan University, Shanghai 200433, China}

\author{Zhe Yuan}
\email{yuanz@fudan.edu.cn}
\affiliation{State Key Laboratory of Surface Physics and Interdisciplinary Center for Theoretical Physics and Information Sciences, Fudan University, Shanghai 200433, China}

\begin{abstract}
We develop a theory for the second-order time-reversal-odd ($\mathcal{T}$-odd) angular-momentum current, incorporating both spin and orbital components. We reveal that besides spin and orbital Berry curvature dipoles, $\mathcal{T}$-odd nonlinear angular-momentum current can originate from disorder-induced mechanisms including coordinate shift, side-jump spin and orbital currents, anomalous scattering amplitude, and skew scattering. A general scaling relation is derived to help distinguish some of these contributions in experiments. Model calculations demonstrate that the orbital component can be comparable to and much larger than the spin component. Our theory lays the groundwork for $\mathcal{T}$-odd nonlinear spin and orbital transport.
\end{abstract}
\maketitle

\renewcommand{\arraystretch}{2}
\emph{Introduction.}{\textemdash}
The importance of disorder scattering in Hall transport of electronic angular momentum is well documented within linear response. For the spin Hall effect, extrinsic mechanisms such as skew scattering and side jump~\cite{gradhand_extSHE_2010,niimi_2011,niimi_2012,zimmermann_2014} can coexist with the intrinsic contribution from spin Berry curvature~\cite{murakami_2003,sinova_SHE_2004,guogy_intSHE_2005,guo_SHE_2008,tanaka_2008} in realistic materials~\cite{sinova_SHE_2015}.
Besides spin, the orbital degree of freedom also carries angular momentum, which has been recognized as equally fundamental~\cite{bernevig_OHE_2005,tanaka_2008,go_2018,choi_OHE_2023,lyalin_2023}. In particular, orbital Hall effect can play an important role in materials already known for efficient spin-current generation~\cite{salemi_2022,go_OHE_2024,mankovsky_2024}, and disorder has also been shown to exert a crucial influence \cite{pezo_2023,liu_ExtrinsicOHE_2024,canonico_2024,tang_vertexOHE_2024,veneri_2025}. 

Meanwhile, nonlinear transport has garnered extensive recent attention owing to its capacity to manifest rich quantum geometric properties and transcend the symmetry limitations inherent in linear transport, such as the nonlinear Hall effect~\cite{gaoY_nonl_in_2014,sodemann_NHE_2015,du_NHE_2018,ma_NAHE_2019,kang_2019,du_NHE2019,xiao_Modified_theory_2019,nandy_2019,wang_intNAHE_2021,liu_intNAHE_2021,gao_NAHE_2023,wang_NAHE_2023,ma_ASkewScatt_2023,huang_scaling_2025,gongzh_NHE_2025}, nonlinear Edelstein effect~\cite{zhou_antidampingSOT_2022,xiao_intNCISP_2022,xiao_ASP_NCISP_2023,guo_extrinsic_2024,kodama_2024a,baek_NEdelstein_2024,oike_2024,oike_2024a,hu_NEdelstein_2025,lu_2025,feng_2025a,niu_2026}, nonlinear spin Hall effect (NSHE)~\cite{yu_NonlinearValleySpin_2014,hamamoto_NSC_2017,hayami_nonlinear_2022,zhang_Intrinsic_2024,wang_Intrinsic_2025} and the recently proposed nonlinear orbital Hall effect (NOHE)~\cite{wanghui_NOHE_2025}. 
For nonlinear Hall effect and nonlinear Edelstein effect, it has been shown that disorder plays an important role~\cite{kang_2019,du_NHE2019,xiao_Modified_theory_2019,nandy_2019,ma_ASkewScatt_2023,huang_scaling_2025,gongzh_NHE_2025,guo_extrinsic_2024}{, with multiple extrinsic contributions such as side jump, skew scattering, spin shift \cite{guo_extrinsic_2024}, and anomalous scattering amplitude \cite{xiao_Modified_theory_2019}. Importantly, characteristic scaling behaviors of nonlinear responses with respect to longitudinal resistivity have been further revealed to provide a useful diagnostic for distinguishing the contributions of different mechanisms~\cite{kang_2019,du_NHE2019,xiao_AHEscaling_2019,guo_extrinsic_2024,huang_scaling_2025,gongzh_NHE_2025}.}
However, a comprehensive understanding of disorder-induced contributions to nonlinear angular-momentum transport{, particularly their microscopic origins and scaling properties,} is still lacking.

In this letter, we develop a theory of time-reversal-odd ($\mathcal{T}$-odd) second-order angular-momentum Hall effect, distinguishing it from $\mathcal{T}$-even contributions~\cite{yu_NonlinearValleySpin_2014,hamamoto_NSC_2017,zhang_Intrinsic_2024,wang_Intrinsic_2025}, and treating spin and orbital contributions on equal footing. We demonstrate several disorder-related mechanisms, including disorder-induced coordinate shift, side-jump angular-momentum current, skew scattering, and anomalous scattering amplitude, in addition to the spin and orbital Berry curvature dipoles.
Furthermore, we propose a scaling law for the $\mathcal{T}$-odd nonlinear angular-momentum Hall effect, offering a practical way to assess the relative importance of these mechanisms in experiments. To quantitatively compare the magnitude of different contributions and the roles of spin and orbital transport, we employ a minimal $\mathcal{PT}$-symmetric model ($\mathcal{P}$ is the inversion symmetry). Our analysis reveals that the $\mathcal{T}$-odd nonlinear angular-momentum Hall effect can be induced by significant disorder effects, and that the orbital component plays a dominant role.

\emph{$\mathcal{T}$-odd second-order angular-momentum Hall effect.}{\textemdash}
The operator of angular-momentum current {flowing along the $\alpha$ direction with angular-momentum polarization along $\beta$} is typically defined as $\hat{j}^{\alpha\beta}_X=-\frac{2e}{\hbar}\{\hat{v}^\alpha,\hat{X}^\beta\}/2$, where $\bm{\hat{X}}=\bm{\hat{L}},\bm{\hat{S}}$ denotes the orbital or spin operator, respectively. The factor $-2e/\hbar$ ensures consistency with the dimension of electric current, {and $e<0$ is the electron charge.} The matrix elements of orbital angular momentum in the crystal momentum representation are given by~\cite{pezo_2022,gobel_2024}
\begin{equation}\label{eq:orbitalAM}
  {\bm{L}}_{n'n}=\frac{ie\hbar^2}{4g_L\mu_B}\sum_{n''\neq n',n}\left(\frac{1}{\varepsilon_{n''n'}}+\frac{1}{\varepsilon_{n''n}}\right){\bm{v}}_{n'n''}\times{\bm{v}}_{n''n},
\end{equation}
where $n$, $n'$, and $n''$ are band indices, $g_L$ is the orbital $g$-factor, and $\mu_B$ is the Bohr magneton. $\varepsilon_{nn'}=\varepsilon_{n\bm{k}}-\varepsilon_{n'\bm{k}}$ is the energy difference between different bands, and $\bm{v}_{nn'}=\langle u_{n\bm{k}}|\partial_{\bm{k}} \hat{H}(\bm{k})/\hbar|u_{n'\bm{k}}\rangle$. Here, the Bloch state is defined as $|l\rangle=\exp(i\bm{k\cdot r})|u_{n\bm{k}}\rangle$ and $l\equiv (n,\bm{k})$. 
Unless otherwise specified, the angular-momentum label $X$ will be omitted in this section, and all analyses are consistent for both types of angular momentum.

Within the semiclassical framework, the total angular-momentum current is expressed as
\begin{equation}\label{eq:totAMC}
  J^{\alpha\beta}=\frac{1}{\mathcal{V}}\sum_l j_{l}^{\alpha\beta}f_{l},
\end{equation}
where $f_l$ is the distribution function for $|l\rangle$, and $\mathcal{V}$ denotes the generalized system volume. $j_{l}$ represents the rank-2 angular-momentum current tensor carried by the perturbed state $|\psi_l\rangle=|l\rangle+|\delta^\text{ext}l\rangle+|\delta^{\bm{E}}l\rangle$, i.e., $j_{l}^{\alpha\beta}=\langle\psi_l|\hat{j}^{\alpha\beta}|\psi_l\rangle$. The corrections $|\delta^\text{ext}l\rangle$ and $|\delta^{\bm{E}}l\rangle$, arising from the disorder and electric field, respectively, modify the expectation of angular-momentum current operator:
\begin{equation}\label{eq:j_ltotal}
  j_{l}^{\alpha\beta}=j_{l}^{\alpha\beta,\text{b}}+j_{l}^{\alpha\beta,\text{sj}}+j_{l}^{\alpha\beta,\text{a}}.
\end{equation}
Here $j_l^\text{b}=\langle l|\hat{j}|l\rangle$ denotes the band angular-momentum current of the unperturbed state and is independent of both the electric field and disorder. 
The second term in Eq.~\eqref{eq:j_ltotal}, originating from disorder-induced correction, represents the angular-momentum current counterpart of the side-jump velocity~\cite{sinitsyn_2006}. This term is referred to as the side-jump angular-momentum current and is proportional to the scattering rate~\cite{xiao_SHEsidejump_2019}, and can be written as
\begin{align}\label{eq:sj}
j_{n\bm k}^{\alpha\beta,\text{sj}}=&-2\pi\sum_{{n}'\bm{k}'}W_{\bm{k}\bm{k'}}\delta(\varepsilon_{{n}\bm{k}}-\varepsilon_{{n}'\bm{k}'}) 
\nonumber\\
&\times\Im\Bigg[\sum_{{n}''\neq{n}'}\frac{\langle u_{{n} \bm{k}}|u_{{n}' \bm{k}'}\rangle\langle u_{{n}'' \bm{k}'}| u_{{n} \bm{k}}\rangle {j}^{\alpha\beta}_{{n}'{n}''}(\bm{k}')}{\varepsilon_{{n}'\bm{k}'}-\varepsilon_{{n}'' \bm{k}'}}
\nonumber\\
&-\sum_{{n}''\neq{n}} \frac{\langle u_{{n}'' \bm{k}}|u_{{n}' \bm{k}'}\rangle \langle u_{{n}' \bm{k}'}|u_{{n} \bm{k}}\rangle {j}^{\alpha\beta}_{{n}{n}''}(\bm{k})}{\varepsilon_{{n} \bm{k}}-\varepsilon_{{n}'' \bm{k}}}\Bigg],
\end{align}
where ${j}^{\alpha\beta}_{{n}{n}'}(\bm{k})=\langle u_{n\bm{k}}|\hat{j}^{\alpha\beta}|u_{n'\bm{k}}\rangle$. $W_{\bm{k}\bm{k'}}=\langle |V_{\bm{kk'}}|^2 \rangle_c$ with $V_{\bm{kk'}}$ as the plane-wave part of disorder-potential matrix elements, and $\langle ...\rangle_c$ stands for the configuration average. 

{To obtain some physical understanding on the side-jump angular-momentum current, one can consider} the specific case of long-range-limit impurity scattering or low-temperature-limit phonon scattering~\cite{xiao_AHEscaling_2019}, where only small angle scattering occurs. In this case, one can expand Eq.~\eqref{eq:sj} to first order in $(\bm{k}-\bm{k}')$ and obtain
\begin{equation}\label{eq:sjexpend}
  j_{n \bm k}^{\alpha\beta,\text{sj}}\approx \sum_\gamma\Omega^{X_\beta}_{n\bm{k},\alpha\gamma} \frac{2\pi}{\hbar} \sum_{\bm{k'}}W_{\bm{kk}'}\delta(\varepsilon_{{n}\bm{k}}-\varepsilon_{{n}\bm{k}'}) (\bm{k}-\bm{k}')_\gamma,
\end{equation}
where $\Omega^{X_\beta}_{n\bm{k},\alpha\gamma}$ is angular-momentum Berry curvature generalized from spin Berry curvature~\cite{guo_SHE_2008,qiao_SHE_2018}, written as
\begin{equation}
  \Omega_{n\bm{k},\alpha\gamma}^{X_\beta}=-2\hbar^2\sum_{n''\neq n}\frac{\Im\left[j^{\alpha\beta}_{nn''}(\bm{k})v^\gamma_{n''n}(\bm{k})\right]}{(\varepsilon_{n\bm{k}}-\varepsilon_{n''\bm{k}})^2}.
\end{equation}

This rank-3 quantum geometric tensor is also manifested in the third term of Eq.~\eqref{eq:j_ltotal}, i.e., 
\begin{equation}\label{eq:j_a}
    j^{\alpha\beta,\text{a}}_l=2\Re\langle l|\hat{j}|\delta^{\bm{E}}l\rangle=-e/\hbar\sum_\gamma\Omega^{X_\beta}_{l,\alpha\gamma}E_\gamma.
\end{equation}
It is analogous to the anomalous velocity induced by an electric field and is referred to as the anomalous angular-momentum current.
Eq.~\eqref{eq:sjexpend}, which links the extrinsic contribution to the intrinsic band geometry, indicates that side-jump angular-momentum current and anomalous angular-momentum current share a similar physical origin: interband virtual transitions induced by external perturbations~\cite{xiao_Modified_theory_2019,guo_extrinsic_2024}.

{
The aforementioned three terms in Eq.~\eqref{eq:j_ltotal} are sufficient for $\mathcal{T}$-odd nonlinear angular-momentum transport. This can be understood as follows: 
The angular-momentum current has opposite time-reversal transformation property to the charge current, thus the physical mechanisms of $\mathcal{T}$-odd nonlinear orbital/spin current responses are parallel to those of the $\mathcal{T}$-even nonlinear charge current response. For the $\mathcal{T}$-even charge current response, it is sufficient to calculate the current expectation value of individual electrons to the leading order of field and disorder-induced corrections, including the Berry curvature anomalous velocity and side-jump velocity \cite{du_NHE2019,xiao_Modified_theory_2019}, while second order corrections to the current expectation value \cite{huang_scaling_2025} are irrelevant.  
In parallel, here the similar calculation procedure applies to the $\mathcal{T}$-odd nonlinear orbital and spin current responses. Indeed, the two correction terms in Eq.~\eqref{eq:j_ltotal} are respectively the angular-momentum counterpart of the Berry curvature anomalous velocity [Eq.~\eqref{eq:j_a}] and of the side-jump velocity \cite{sinitsyn_2006,xiao_2017} [Eq.~\eqref{eq:sj}]. }

\begin{table*}[tpb]
  \centering
  \caption{Second-order Angular-momentum Hall Conductivities of the tilted four-band Dirac model in Eq.~\eqref{eq:model}. Here, $\chi_{xyy}^{S_z}$ and $\chi_{xyy}^{L_z}$ denote the spin and orbital contributions, respectively.}
  \label{tab:AMHE}
  \begin{tabular*}{2\columnwidth}{@{\extracolsep{\fill}}ccc@{}}
  \hline\hline
  Mechanism & $\chi_{xyy}^{S_z}$ & $\chi_{xyy}^{L_z}$ \\
  \hline
  Angular-momentum Berry Curvature Dipole & $-\frac{3 \Delta  {e}^3 v^2 w \left({\varepsilon_F^2}-\Delta ^2\right)}{\pi  {\varepsilon_F^3} n_{i}{V_0}^2 \hbar  \left(3 \Delta ^2+{\varepsilon_F^2}\right)}$ & $\frac{5 \Delta ^2 {e}^4 v^4 w \left(\Delta ^2-{\varepsilon_F^2}\right)}{\pi  {\varepsilon_F^5} {g_L} {\mu_B} {n_i}{V_0^2} \hbar ^2 \left(3 \Delta ^2+{\varepsilon_F^2}\right)}$\\
  Side-jump Angular-momentum Current &
  $-\frac{\Delta  {e}^3 v^2 w \left(5 {\varepsilon_F^2}-33 \Delta ^2\right) \left({\varepsilon_F^2}-\Delta ^2\right)}{2 \pi  {\varepsilon_F^3} n_{i}{V_0}^2 \hbar  \left(3 \Delta ^2+{\varepsilon_F^2}\right)^2}$ &
  $\frac{3 \Delta ^2 {e}^4 v^4 w \left({\varepsilon_F^2}-\Delta ^2\right) \left(23 \Delta ^2-11{\varepsilon_F^2}\right)}{2 \pi  {\varepsilon_F^5} {g_L} {\mu_B} {n_i}{V_0^2} \hbar ^2 \left(3 \Delta ^2+{\varepsilon_F^2}\right)^2}$\\
  Coordinate Shift &
  $\frac{\Delta  {e}^3 v^2 w \left(\varepsilon_F^2-\Delta^2\right) \left(17 \Delta ^2+3 \varepsilon_F^2\right)}{2 \pi  {\varepsilon_F^3} n_{i}{V_0}^2 \hbar  \left(3 \Delta ^2+{\varepsilon_F^2}\right)^2}$ &
  $-\frac{3 \Delta ^2 {e}^4 v^4 w \left(7 {\varepsilon_F^2}-19 \Delta ^2\right) \left({\varepsilon_F^2}-\Delta ^2\right)}{2 \pi  {\varepsilon_F^5} {g_L} {\mu_B} {n_i}{V_0^2} \hbar ^2 \left(3 \Delta ^2+{\varepsilon_F^2}\right)^2}$\\
  Anomalous Scattering Amplitude&
  $-\frac{3 \Delta  {e}^3 v^2 w \left({\varepsilon_F^2}-\Delta ^2\right)^2}{ \pi  {\varepsilon_F^3} n_{i}{V_0}^2 \hbar  \left(3 \Delta ^2+{\varepsilon_F^2}\right)^2}$ 
  &
  $-\frac{3 {e}^4 v^4 w \Delta^2\left(\Delta ^2-  {\varepsilon_F^2}\right)^2}{\pi  {\varepsilon_F^5} {g_L} {\mu_B} {n_i}{V_0^2} \hbar ^2 \left(3 \Delta ^2+{\varepsilon_F^2}\right)^2}$\\
  Conventional Skew Scattering&
  $\frac{2\Delta  {e}^3 v^2 V_1^3 w \left({\varepsilon_F^2}-\Delta ^2\right)^2 \left(9 \Delta ^2+5 {\varepsilon_F^2}\right)}{\pi  {\varepsilon_F^2} n_i^2V_0^6 \hbar  \left(3 \Delta ^2+{\varepsilon_F^2}\right)^3}$ &
  $-\frac{6 \Delta ^2 {e}^4 v^4 {V_1^3} w \left({\varepsilon_F^2}-7 \Delta ^2\right) \left({\varepsilon_F^2}-\Delta ^2\right)^2}{\pi  {\varepsilon_F^4} {g_L} {\mu_B} {n_i^2}{V_0^6} \hbar ^2 \left(3 \Delta ^2+{\varepsilon_F^2}\right)^3}$\\
  Gaussian Skew Scattering&
  $\frac{\Delta  {e}^3 v^2 w \left({\varepsilon_F^2}-\Delta ^2\right)^2 \left(77 \Delta ^2+13 {\varepsilon_F^2}\right)}{2 \pi {\varepsilon_F^3} n_{i}{V_0}^2 \hbar  \left({\varepsilon_F^2}+3 \Delta ^2 \right)^3}$ &
  $\frac{3 \Delta^2 {e}^4 v^4 w \left(51 \Delta ^2-13{\varepsilon_F^2}\right) \left(\Delta ^2-  {\varepsilon_F^2}\right)^2}{2 \pi {\varepsilon_F^5} {g_L} {\mu_B} {n_i}{V_0^2} \hbar ^2 \left({\varepsilon_F^2}+3 \Delta ^2 \right)^3}$ \\
  \hline \hline
  \end{tabular*}
\end{table*}

The distribution function $f_l$ of electron states is described by the semiclassical Boltzmann equation. In a homogeneous system, it satisfies
\begin{equation}\label{eq:Boltz}
  \frac{e}{\hbar}\bm{E}\cdot\frac{\partial f_l }{\partial {\bm{k}}}=-\sum_{l'}\left(\omega_{l'l}f_{l}-\omega_{ll'}f_{l'}\right),
\end{equation}
where $\omega_{ll'}$ is the scattering rate from $l'$ to $l$. As discussed in previous studies~\cite{du_NHE2019,xiao_Modified_theory_2019, guo_extrinsic_2024}, $f_l$ can be expanded as $f_l\approx \sum_m f_{m,l}$, with $f_{m,l}$ representing the $m$th-order correction in the electric field $\bm{E}$. In addition, $f_l$ can be decomposed according to distinct scattering contributions~\cite{sinitsyn_AHE_2007}, {as detailed in the Supplemental Material~\cite{supp}}:
\begin{align}\label{eq:DF}
  f_{m,l}=&f_{m,l}^{\text{D}}+f_{m,l}^{\text{cs}}+f_{m,l}^{\text{asa}}+f_{m,l}^{\text{csk}}+f_{m,l}^{\text{Gsk}}.
\end{align}
Here, $f_{m,l}^\text{D}$ arises from the symmetric second-order scattering, while $f_{m,l}^\text{csk}$ and $f_{m,l}^\text{Gsk}$ originate from the antisymmetric parts of the third- and fourth-order scattering amplitudes, respectively, commonly referred to as conventional and Gaussian skew scattering (SK). $f_{m,l}^{\mathrm{cs}}$ and $f_{m,l}^{\mathrm{asa}}$ arise from the field-induced coordinate shift and anomalous scattering amplitude, respectively~\cite{xiao_Modified_theory_2019}.

Defining the $\mathcal{T}$-odd second-order angular-momentum current conductivity as $J^{\alpha\beta}=\chi^{X_\beta}_{\alpha\gamma\delta}E_{\gamma}E_\delta$ and comparing with Eq.~\eqref{eq:totAMC}, only the following terms correspond to $\mathcal{T}$-odd mechanisms:
\begin{align}\label{eq:chi}
  \chi^{X_\beta,\text{tot}}_{\alpha\gamma\delta}=&\chi^{X_\beta,\text{ABD}}_{\alpha\gamma\delta}+\chi^{X_\beta,\text{SJC}}_{\alpha\gamma\delta}+\chi^{X_\beta,\text{CS}}_{\alpha\gamma\delta}
  \nonumber\\
  &+\chi^{X_\beta,\text{ASA}}_{\alpha\gamma\delta}+\chi^{X_\beta,\text{SK}}_{\alpha\gamma\delta}.
\end{align}
Here, the first term in the right-hand side originates from the product of the anomalous angular-momentum current and the Drude Fermi-surface shift $j_l^\text{a}f^\text{D}_{1,l}$, which can be rewritten in the form of the angular-momentum Berry curvature dipole (ABD) in momentum space. The second term arises from higher-order Drude shift with side-jump angular-momentum current (SJC). The remaining three terms combine the unperturbed band current $j_l^{\text{b}}$ with different components of $f_{2,l}$: $j_l^{\text{b}}f_{2,l}^{\text{cs}}$ for coordinate shift (CS), $j_l^{\text{b}}f_{2,l}^{\text{asa}}$ for anomalous scattering amplitude (ASA), and $j_l^{\text{b}}(f_{2,l}^{\text{csk}}+f_{2,l}^{\text{Gsk}})$ for skew scattering. The summation over $l$ has been omitted for brevity. The last four terms are new disorder-induced contributions that have not been discussed in previous works on $\mathcal{T}$-odd nonlinear angular-momentum current~\cite{hayami_nonlinear_2022,zhang_Intrinsic_2024}.
All other combinations not listed here are $\mathcal{T}$-even. For example, the nonlinear Drude term $j_l^\text{b} f_{2,l}^{\text{D}}$, which possesses the same symmetry as the intrinsic contribution~\cite{zhang_Intrinsic_2024,wang_Intrinsic_2025}, is excluded from the present consideration.
{The detailed time-reversal properties of the microscopic quantities appearing in Eqs.~\eqref{eq:j_ltotal} and \eqref{eq:DF}, as well as those of their combinations, are detailed in the Supplemental Material~\cite{supp}.}

\begin{figure*}[t]
  \centering
  \includegraphics[width=2\columnwidth]{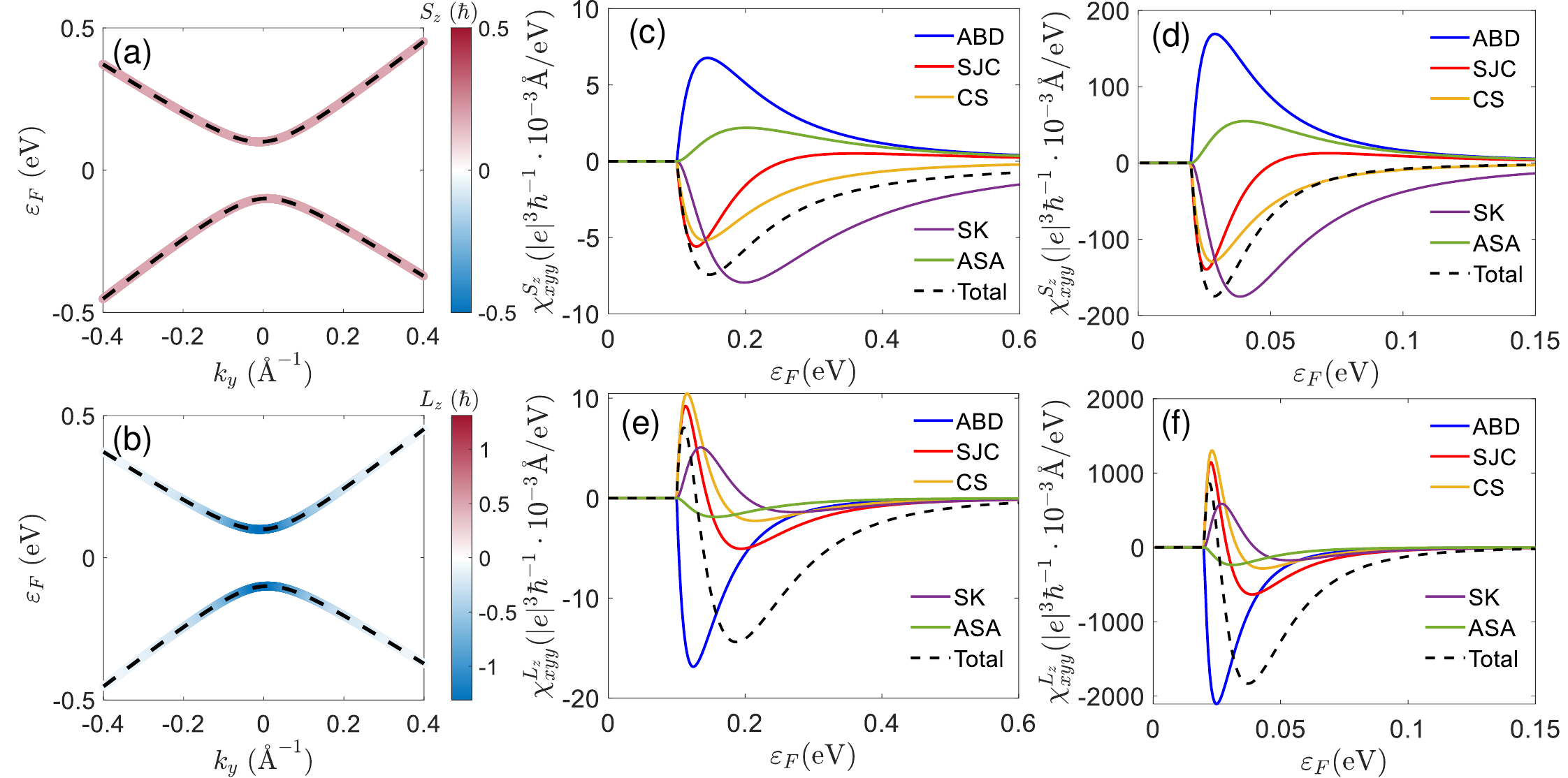}\caption{
  Results obtained from the tilted four-band Dirac model in Eq.~\eqref{eq:model}. {The upper (lower) panels correspond to the spin (orbital) case.}
  (a) and (b) show the energy bands along $k_y$ with spin and orbital, respectively. Dashed lines denote doubly degenerate bands, with the color bar indicating the angular momentum of the spin-up block; the spin and orbital components of the spin-down block have opposite signs and are omitted for clarity.
  (c) and (d) present the NSHE response and (e) and (f) show the NOHE response, as a function of Fermi energy. $\Delta$ in (a-c,e) is set to $0.1$ eV, {whereas a smaller value, $\Delta=0.02$ eV, is used in (d,f) to highlight the enhancement of the response associated with a reduced band gap}.
  The other parameters are set to $w=0.1$ eV$\cdot$\AA, $v=1$ eV$\cdot$\AA, $n_i V_0^2=10^2$ (eV$\cdot$\AA)$^2$, $n_i V_1^3=10^4$ eV$^3\cdot$\AA$^4$, and $g_L=1$. }\label{fig:ef}
\end{figure*}

\emph{Model results.}{\textemdash}
To gain deeper insight into the $\mathcal{T}$-odd angular-momentum current and compare the nonlinear spin and orbital Hall effects, we employ the tilted four-band Dirac model~\cite{ma_ASkewScatt_2023,huang_scaling_2025}, which breaks both $\mathcal{P}$ and $\mathcal{T}$ symmetries while preserving $\mathcal{PT}$ symmetry:
\begin{equation}\label{eq:model}
H(\bm{k})=wk_y+v(k_x\sigma_x+k_y\sigma_y)+\Delta\sigma_z s_z,
\end{equation}
where $\bm{k}=(k_x,k_y)$ is the wave vector, and $s_i$ and $\sigma_i$ are Pauli matrices representing spin and orbital spaces, respectively. The parameters $w$, $v$, and $\Delta$ control the tilt, velocity, and gap of the model, while the last term couples spin and orbital degrees of freedom. 
The Hamiltonian is block-diagonal, with $s_z$ being a good quantum number that takes opposite signs in different blocks. The bands remain doubly degenerate owing to $\mathcal{PT}$-symmetry~\cite{tangpz_2016}. For the two-dimensional model in Eq.~\eqref{eq:model}, only the $z$ component of orbital angular momentum survives, given by ${L}^{z,\pm} = \pm \sigma_0 (e/g_L \mu_B)(v^2 \Delta / 2 \varepsilon_0^2)$, where $\varepsilon_0 = \sqrt{\Delta^2 + v^2k^2}$ is the band energy in the absence of tilt and $\pm$ correspond to the spin-resolved blocks. One can find that $|L^z|$ decreases as the Fermi energy moves away from the gap, while the spin remains unchanged, as illustrated in Figs.~\ref{fig:ef}(a) and (b). The responses for NSHE and NOHE are shown in Fig.~\ref{fig:ef}(c-f). Here we consider short-range random scalar impurities scattering with the potential $V(\bm{r}) = \sum_i V_i \delta(\bm r - \bm R_i)$. The disorder averages are specified as $\langle V_i \rangle_c = 0$, $\langle V_i^2 \rangle_c = V_0^2$, and $\langle V_i^3 \rangle_c = V_1^3$. To obtain analytical results, we focus on the limit $w \ll v$ and retain only terms up to first order in $w$. In addition, the isotropic constant-relaxation-time approximation is adopted~\cite{du_NHE2019,xiao_Modified_theory_2019}. The explicit expressions are summarized in Tables~\ref{tab:AMHE}.

Figs.~\ref{fig:ef}(c-f) show that the orbital Hall effect is non-negligible compared with the spin Hall effect in the nonlinear regime, analogous to what is observed in linear transport for many materials~\cite{go_OHE_2024,mankovsky_2024,abrao_AOHE_2025,costa_OHE_2025}. All components of the conductivity vanish for $|\varepsilon_F| < \Delta$, indicating that the mechanisms considered here are governed by Fermi-surface transport.
Furthermore, the orbital Hall conductivity decays more rapidly than the spin Hall conductivity with increasing $\varepsilon_F$, reflecting the suppression of the orbital angular momentum magnitude $|L_z|$. As shown in Figs.~\ref{fig:ef}(c–f), all disorder-induced mechanisms contribute appreciably in the relevant energy window. Upon reducing the band gap [Figs.~\ref{fig:ef}(d,f)], both the NSHE and NOHE are enhanced. Notably, the orbital contribution progressively dominates the total angular-momentum Hall response in the vicinity of the gap.

To elucidate this behavior, we examine the ratio of the two angular momentum conductivities arising from a given mechanism at $\varepsilon_F \approx \Delta$. Taking the angular-momentum Berry curvature dipole contribution as a representative example, the ratio is estimated from Tables~\ref{tab:AMHE} as
\begin{equation}
\frac{\chi_{xyy}^{L_z,\mathrm{ABD}}}{\chi_{xyy}^{S_z,\mathrm{ABD}}}
\approx \frac{5 e v^2}{3 g_L \mu_B \hbar}\frac{1}{\Delta}.
\end{equation}
A similar $\Delta^{-1}$ scaling is found for the other disorder-induced mechanisms. This scaling demonstrates that the NOHE can substantially exceed the NSHE in small-gap systems, underscoring the crucial role of orbital angular momentum in nonlinear angular momentum transport. {Importantly, this finding provides a systematic framework to predict and quantify orbital contributions in realistic materials. In particular, the pronounced $\Delta^{-1}$ scaling suggests that small-gap systems are promising platforms for experimentally probing and utilizing the nonlinear orbital Hall response, offering new opportunities for controlling angular-momentum currents beyond conventional spin-based mechanisms.}

\emph{Discussion.}{\textemdash}
Although spin and orbital angular momentum currents both contribute to the total angular momentum transport as shown above, their symmetries are fundamentally distinct when viewed from the perspective of spin groups. In the absence of SOC, the symmetry of the system is allowed to be described by spin group~\cite{smejkal_spingroup_2022}. For collinear magnetic structures, the $\mathcal{T}$-odd nonlinear spin Hall effect can exist even without SOC—arising purely from the staggered exchange fields of opposite sublattices~\cite{hayami_nonlinear_2022}, while the $\mathcal{T}$-odd nonlinear orbital Hall effect requires spin-orbit coupling. 

{Our model calculations and the discussion of symmetries in Supplemental Material~\cite{supp} suggest that $\mathcal{PT}$-symmetric antiferromagnets with topological band structures, such as CuMnAs and even-layer $\mathrm{MnBi_2Te_4}$, are promising candidates for probing the $\mathcal{T}$-odd NSHE and NOHE, where the corresponding $\mathcal{T}$-even effects are forbidden. Experimental studies on these materials may help deepen the understanding of $\mathcal{T}$-odd nonlinear angular momentum transport.}

To understand the disorder dependence of nonlinear angular-momentum current, it is essential to introduce the scaling law, which has been widely applied in the analysis of electric transport experiments~\cite{tian_scaling_2009,hou_scaling_2015,ma_NAHE_2019,gao_NAHE_2023}. In real systems, both static disorder, such as defects and impurities, and dynamic disorder, such as phonons, are usually present. Here, we neglect the interplay between different types of disorder, so that Matthiessen's rule applies: $\rho = \sum_i \rho_i$, where $\rho_i$ denotes the resistivity induced by the $i$-th type of disorder, and $\rho$ is the total longitudinal resistivity. The general scaling form of the second-order angular-momentum current can then be written as
\begin{align}\label{eq:scaling}
  \chi=\frac{C}{\rho}+\sum_{i\in S}{A_i}\frac{\rho_i}{\rho^3}+\sum_i C_i\frac{\rho_i}{\rho^2}+\sum_{i,j}C_{ij}\frac{\rho_i\rho_j}{\rho^3}.
\end{align}
Here, $A_i$ and $C_{ij}$ correspond to the conventional and Gaussian skew-scattering terms, respectively, while $C_i$ arises from the side-jump angular-momentum current, coordinate shift, and anomalous scattering amplitude, which share the same scaling behavior, i.e., $C_i=C_i^\mathrm{SJC}+C_i^\mathrm{CS}+C_i^\mathrm{ASA}$. The parameter $C$ represents the ABD term. {$i\in S$ indicates that only static disorder is taken into account.} 

For illustration, we restrict ourselves to two types of scattering: impurity scattering ($\rho_0$) and phonon scattering ($\rho_1$). In this case, Eq.~\eqref{eq:scaling} can be rewritten as
\begin{equation}\label{eq:scaling_new}
    \chi=A_0 {\rho_0}/{\rho^3}+B/\rho+B'\rho_0/\rho^2+B''\rho_0^2/\rho^3,
\end{equation}
where $B=C+C_1+C_{11}$, $B'=C_0-C_1+C_{01}+C_{10}-2C_{11}$, and $B''=C_{00}+C_{11}-C_{01}-C_{10}$. $A_0$ and $\rho_0$ can be determined from the low-temperature limit. The coefficients $B$, $B'$ and $B''$ can then be extracted by fitting the temperature dependence of $\chi$. Notably, the terms proportional to $B'$ and $B''$ originate purely from extrinsic mechanisms.
{
At low temperatures, $\rho \to \rho_0$ and the scaling law reduces to $\chi=A_0/ {\rho_0^2}+(C+C_0+C_{00})/\rho_0$, thus the conventional skew scattering contribution is anticipated to be dominant in highly conductive systems.}
{In real materials, it remains challenging to quantitatively calculate the disorder-induced contributions to nonlinear effects. However, the above scaling-law analysis provides a practical route for experimentally evaluating these disorder-induced effects.
}

In summary, we have developed a semiclassical theory for disorder-induced $\mathcal{T}$-odd nonlinear angular-momentum Hall effect involving both spin and orbital components, and proposed a scaling law for these effects. Our analysis further shows that the orbital contribution plays a crucial role comparable to the spin one. These results establish the fundamental framework of $\mathcal{T}$-odd nonlinear angular momentum transport.

\emph{Acknowledgments.}{\textemdash}
This work was supported by the National Natural Science Foundation of China (Grants No. 12574115 and No. 12374101). C.X. was sponsored by National Natural Science Foundation of China (Grant No. 12574114) and the start-up funding from Fudan University.

\twocolumngrid

\bibliography{ref}

\end{document}